\documentclass{article}
\usepackage{latexsym}
\usepackage{amsmath}
\usepackage{epsfig}
\usepackage{epsf}
\usepackage{latexsym}
\usepackage{url}
\usepackage{xspace}
\usepackage{cite}

\advance \topmargin by -\headheight
\advance \topmargin by -\headsep
\evensidemargin \oddsidemargin
\newcommand{\old}[1]{
}

\newcommand{\full}[1]{#1}              

\newcommand{\short}[1]{}               

\newcommand{\comment}[1]%
{%
  \mbox{}\marginpar{\textsl{\fbox{\vbox{\tiny\raggedright\hspace{0pt}#1}}}}%
}%

\newtheorem{lemma}{Lemma}

\newtheorem{corollary}[lemma]{Corollary}
\newtheorem{theorem}[lemma]{Theorem}

\newtheorem{proposition}[lemma]{Proposition}
\newtheorem{conj}[lemma]{Conjecture}

\makeatletter
\def\GrabProofArgument[#1]{ (#1): \egroup\ignorespaces}
\def\proof{\noindent\textbf\bgroup Proof%
           \@ifnextchar[{\GrabProofArgument}{: \egroup\ignorespaces}}

\makeatother

\newlength{\saveparindent}
\newlength{\saveparskip}
\newenvironment{newitemize-a}{%
\begin{list}{$\bullet$}{\labelwidth=19pt%
\labelsep=7pt \leftmargin=18pt \topsep=3pt%
\setlength{\listparindent}{\saveparindent}%
\setlength{\parsep}{\saveparskip}%
\setlength{\itemsep}{3pt} }}{\end{list}}

\newenvironment{specialitemize}{%
\begin{list}{$\bullet$}{\labelwidth=19pt%
\labelsep=7pt \leftmargin=25pt \topsep=3pt%
\setlength{\listparindent}{\saveparindent}%
\setlength{\parsep}{\saveparskip}%
\setlength{\itemsep}{3pt} }}{\end{list}}

\begin{document}

\title{The Freeze-Tag Problem: How to Wake Up a Swarm of
Robots\thanks{An extended abstract was presented at
the {\em 13th Annual ACM-SIAM Symposium on Discrete Algorithms (SODA)\/},
San Francisco, January 2002~\cite{abfms-ftp-02}.}}
\author{
  Esther M. Arkin\footnotemark[2]%
\and
  Michael A. Bender\footnotemark[3]
\and
  S\'andor P. Fekete\footnotemark[4]%
\and
  Joseph S. B. Mitchell\footnotemark[2]
\and
  Martin Skutella\footnotemark[5]
}
\date{}

\maketitle

\renewcommand{\thefootnote}{\fnsymbol{footnote}}

\footnotetext[2]{
    Department of Applied Mathematics and Statistics,
    State University of New York, Stony Brook, NY 11794-3600, USA.
    Email:~\{\texttt{estie}, \texttt{jsbm}\}\texttt{@ams.sunysb.edu}.}
\footnotetext[3]{
    Department of Computer Science, State University of New York,
    Stony Brook, NY 11794-4400, USA.
    Email:~\texttt{bender@cs.sunysb.edu}.}
\footnotetext[4]{
    Department of Mathematical Optimization,
    Braunschweig University of Technology,
    Pockelsstr.~14, 38106 Braunschweig, Germany.
    Email:~\texttt{s.fekete@tu-bs.de}.}
\footnotetext[5]{
    Fachbereich Mathematik, Universit\"at Dortmund, 44221 Dortmund,
    Germany.
    Email:~\texttt{martin.skutella@uni-dortmund.de}.}

\renewcommand{\thefootnote}{\arabic{footnote}}

\begin{abstract}
An optimization problem that naturally arises in the study of
swarm robotics is the {\em Freeze-Tag Problem (FTP)\/}
of how to awaken a set of ``asleep'' robots, 
by having an awakened robot move to their locations. 
Once a robot is awake, it can assist in awakening other slumbering robots.
The objective is to have all robots awake as early as possible.
While the FTP bears some resemblance to problems from
areas in combinatorial optimization such as routing, broadcasting,
scheduling, and covering, its algorithmic characteristics are
surprisingly different.  

We consider both scenarios on graphs and in geometric environments.
In graphs,
robots sleep at vertices and there is a length function on the
edges. Awake robots travel along edges, with time 
depending on edge length. 
For most scenarios, we consider 
the offline version of the problem,
in which each awake robot knows the position of all other robots. 
We prove that the problem is NP-hard, even
for the special case of star graphs.  We also establish hardness of
approximation, showing that it is NP-hard to obtain an approximation
factor better than $5/3 $, even for graphs of bounded degree.

These lower bounds are complemented with several positive algorithmic
results, including:

\begin{itemize}
\item We show that the natural greedy strategy on star graphs
has a tight worst-case performance of $ 7/3$ and
give a polynomial-time approximation scheme (PTAS)
for star graphs.

\item We give a simple $O(\log\Delta)$-competitive online algorithm
for graphs with maximum degree $\Delta$ and locally bounded edge weights.


\item We give a PTAS, running in nearly linear time, for geometrically
embedded instances.
\end{itemize}
\end{abstract}


{\bf AMS-Classification:}
68Q25, 68T40, 68W25, 68W40, 90B35.

\pagestyle{myheadings}
\thispagestyle{plain}
\markboth{E. M. ARKIN ET AL.}{Freeze-Tag: Waking up a swarm of robots}

\section{Introduction}

The following problem naturally arises in the study of {\em swarm
robotics\/}.  Consider a set of $n$ {\em robots\/}, modeled as points in
some metric space (e.g., vertices of an edge-weighted graph).
Initially, there is one {\em awake} or {\em active}
robot and all other robots are {\em
asleep\/}, that is, in  a {\em stand-by mode\/}.
Our objective is to ``wake up'' all of
the robots as quickly as possible.
In order for an active robot to awaken a sleeping robot,
the awake robot must travel to the location of the slumbering robot.
Once awake, this new robot is
available to assist in rousing other robots.
The objective is to minimize the {\em makespan\/}, that is, the
time when the last robot awakens.

This awakening problem is reminiscent of the children's game of ``freeze-tag'',
in which the person who is ``it'' tags other players to ``freeze'' them.
A player remains ``frozen'' until
an unfrozen player (who is not ``it'')
rescues the frozen player by tagging him and thus
unfreezing him.
Our problem arises when
there are a large number $n$ of frozen players, and
one (not ``it'') unfrozen player, whose goal it is to
unfreeze
the rest of the players as quickly as possible.
(We do not
take into consideration
the effect of the person who is ``it'', who is likely running around
and re-freezing the players that become defrosted!)
As soon
as a player becomes unfrozen, he is available to assist in helping
other frozen players, so there is a cascading effect.  Due to the
similarity with this child's game, we dub our problem the
{\em Freeze-Tag Problem (FTP)\/}.

Other applications of the FTP arise in the context of
distributing data (or some other commodity), where physical proximity
is required for transmittal.
Proximity may be required because wireless
communication is too costly in terms of bandwidth or
because there is too much of a
security risk.
Solutions to the FTP determine how to
propagate the data to the entire set of
participants in the most efficient manner.

In this paper we introduce and present algorithmic results
for the FTP, a problem that
 arises naturally as a hybrid of problems from the
areas of broadcasting, routing, scheduling, and network design.
We focus on the offline version of the problem, in which each 
awake robot knows the position of all other robots, and is able
to coordinate its moves with the other robots.
The FTP is a {\em network design\/} problem because
the optimal schedule is determined by a spanning
binary tree of minimum depth in a (complete) weighted graph.
As in {\em broadcasting
problems\/}, the goal is to disseminate information in a network.  The FTP
has elements of optimal {\em routing\/},
because robots must travel to awaken
others or to pass off information.
The FTP can even be thought of as a parallel version of the
traveling salesmen problem,
in which salesmen are posted in each city.
Finally, the FTP has elements of
{\em scheduling\/}
(where the number of processors increases over time),
and scheduling techniques (e.g., use of min-sum criteria)
are often relevant.
Finally  we note that
given the practical motivation of the problem (e.g., in robotics),
there is
interest in considering {\em online versions\/}, where each robot can
only see its immediate neighborhood in the graph.

\paragraph{Related Work.}
There is an abundance of prior work on the dissemination of data in a
graph. Most closely related to the FTP are the {\em minimum broadcast
time problem\/}, the {\em multicast problem\/}, and the related {\em
minimum gossip time problem\/}.  See~\cite{Hed88a} for a survey;
see~\cite{STOC98*448,focs94*202} for approximation results.
However, the proximity required in the FTP leads to significant
differences: While the broadcast problem can be solved in polynomial
time in tree networks, the FTP turns out to be NP-hard on the
seemingly easy class of weighted stars.

In the field of robotics, several related algorithmic problems have
been studied for controlling swarms of robots to perform various
tasks, including environment
exploration~\cite{albers97exploring,%
albers99exploring,bruckstein97probabilistic,gage00navmap,icking-exploring,wagner96cooperative,wagner-mac},
robot
formation~\cite{bonabeau-book,sugihara96distributed,suzuki99distributed},
searching~\cite{wagner98efficiently}, and
recruitment~\cite{wagner95cooperative}.  Ant behaviors have inspired
algorithms for multi-agent problems such as searching and covering;
see, e.g.,
\cite{wagner96cooperative,wagner98efficiently,wagner95cooperative}.
Multi-robot formation in continuous and grid environments has been
studied recently by Sugihara, Suzuki, Yamashita, and Dumitrescu; see
\cite{suzuki99distributed,sugihara96distributed,DumitrescuSuzukiYamashita2002}.
The objective is for distributed robots to form shapes such as circles
of a given diameter, lines, etc. without using global control.
Teambots, developed by Balch~\cite{teambots} in Java, is a popular
general-purpose multi-robot simulator used in studying swarms of
robots. \cite{habfm-ardrue-03} and the video \cite{habfm-odsr-03}
deal with the distributed, online problem of dispersing a swarm
of robots in an unknown environment.

Gage~\cite{gage98evolutionary,gage92sensor,dwgage-website}
has proposed the development of command and control tools
for arbitrarily large swarms of microrobots.  He originally posed
to us the problem of how to ``turn on'' a large swarm of robots
efficiently; this question is modeled here as the FTP.

Another related problem is to consider variants where all robots are mobile,
but they still have to meet in order to distribute important information.
The two-robot scenario with initial positions unknown to both
players is the problem of rendezvous search that has received
quite a bit of attention, see~\cite{AndFekRend,roy-collaborative}
and the relatively recent book by Alpern and Gal~\cite{AlpernGal} for an overview.

In subsequent work on the FTP, Sztainberg, Arkin, Bender, and
Mitchell~\cite{sabm-teahf-04}
have analyzed and implemented heuristics for the FTP. They showed that the
greedy strategy gives a tight approximation bound of
$\Theta(\sqrt{\log n})$ for the case of points in the plane and,
more generally, $\Theta((\log n)^{1-1/d})$ for points in $d$
dimensions. They also presented experimental results on classes of
randomly generated data, as well as on data sets from the TSPLIB
repository~\cite{TSPLIB-ref}.

Arkin, Bender, Ge, He, and Mitchell~\cite{ABGHM-03} gave an
$O(1)$-approximation algorithm for the FTP in unweighted graphs,
in which there is one asleep robot at each node, and they showed
that this version of the FTP is NP-hard. They generalized to the
case of multiple robots at each node; for unweighted edges,
they obtained a $\Theta(\sqrt{\log n})$ approximation, and for
weighted edges, they obtained an $O((L/d)\log
n+1)$-approximation algorithm, where $L$ is the length of the
longest edge and $d$ is the diameter of the graph.

More recently, K\"{o}nemann, Levin, and Sinha~\cite{KonemannLeSi03}
gave an $O(\sqrt{\log n})$-approximation algorithm for the general
FTP, in the context of the \emph{bounded-degree minimum diameter
spanning tree problem}.  Thus, the authors answer in the
affirmative an important open question
from~\cite{abfms-ftp-02,sabm-teahf-04,TSPLIB-ref}.
In contrast to the results from~\cite{KonemannLeSi03}, our paper
gives tighter approximation bounds but for particular versions of
the FTP.

\paragraph{Intuition.}
An algorithmic dilemma of the
FTP is the following:  A robot must decide
whether or not to awaken a small nearby cluster to obtain a modest number of helpers
quickly
or whether to awaken a distant but populous cluster to obtain many helpers,
but after a longer delay.
This dilemma is compounded because
clusters may have uneven densities
so that clusters may be within clusters.
Even in the simplest cases,
packing and partitioning problems are embedded in the FTP;
thus the FTP on stars is NP-hard because of inherent partitioning problems.
What makes the Freeze-Tag Problem particularly intriguing is that
while it is fairly
straightforward to obtain an algorithm
that is $O(\log n)$-competitive for the FTP with locally bounded
edge weights,
it is highly nontrivial to
obtain an $o(\log n)$ approximation bound
for general metric spaces, or  even on
special graphs such as trees.

Some of our results are specific to {\em star metrics\/},
which arise as an important
tool in obtaining approximation
algorithms in more general metric spaces, as shown, e.g., in
\cite{bartal98approximating,charikar98approximating,konjevod97on}.
(See our conference version \cite{abfms-ftp-02} for further results
on a generalization called {\em ultrametrics\/}.)
We also study a geometric variant of the problem in which the robots are
located at points of a geometric space and travel times are given by
geometric distances.

\subsection{Summary of Results}
\label{subsec:results}

This paper presents the following results:

\begin{specialitemize}

\item
  We prove that the Freeze-Tag Problem is NP-hard, even for the
  case of star graphs with an equal number of robots at each vertex 
  (Section 2.3).
  Moreover, there exists a polynomial-time
  approximation scheme (PTAS) for this case (section 2.4.
 We analyze the greedy
  heuristic, establishing a tight performance bound of 7/3 (Section 2.2).
We show an $O (1) $-approximation algorithm
for more general star graphs that can have clusters of robots
at the end of each spoke (Section 2.5).

\item
   We give a simple linear-time online algorithm
  that is $O(\log\Delta)$-competitive
  for the case of general weighted graphs of maximum
  degree $\Delta$ that have ``locally bounded'' edge weights, meaning that
the ratio of the largest to the smallest edge weight
among edges incident to a vertex is bounded (Section 3.1).
  On the other hand, we show for the offline problem
  that finding a solution within an approximation factor
  less than $5/3$ is NP-hard, even for graphs of maximum degree~$5$ (Section 3.2).

\item
  We give a PTAS for geometric instances of the FTP in any
  fixed dimension, with distances given by an $L_p$ metric.
  Our algorithm runs in near-linear time, $O(n\log
  n+2^{poly(1/\varepsilon)})$, with the nonpolynomial dependence on
  $\varepsilon$ showing up only as an additive term in the time
  complexity (Section 4).
\end{specialitemize}

See Table~\ref{tab:overview} for an overview of results for the FTP.


\begin{table}
\small{
\centerline{
\begin{tabular}{|l l|c|c|c|}
\hline
Version & Variant &  Complexity & Approx. Factor & LB for Factor \\
\hline \hline
General graphs & weighted & NPc (Sec. 2.3)&  $O(\sqrt{\log n})$ \cite{KonemannLeSi03} & 5/3 (Sec. 3.2) \\
             & unweighted & NPc \cite{sabm-teahf-04} & $O(\sqrt{\log n})$ \cite{sabm-teahf-04} & {\em open} \\
\hline
Trees & weighted & NPc (Sec. 2.3) & $O(\sqrt{\log n})$ \cite{KonemannLeSi03} & {\em open} \\
      & unweighted & NPc (Sec. 2.3) & $O(\sqrt{\log n})$ \cite{sabm-teahf-04} & {\em open} \\
\hline
Ultrametrics & weighted & NPc (Sec. 2.3) & $2^{O(\sqrt{\log\log n})}$ \cite{abfms-ftp-02} & {\em open} \\
\hline
Stars & weighted, $\rho(v) \equiv c$, greedy & NPc (Sec. 2.3) & 7/3 (Sec. 2.2) & 7/3 (Sec. 2.2) \\
Stars & weighted, $\rho(v) \equiv c$ & NPc (Sec. 2.3) & $1+\varepsilon$ (Sec. 2.4) & n/a \\
& weighted, $\rho(v)$ arbitrary & NPc (Sec. 2.3) & 14 (Sec.2.5) & {\em open} (Conj.\ref{con:uneven})\\
& unweighted, $\rho(v)$ & P (Sec. 2) & n/a & n/a \\
\hline
\hline
Geometric & $L_p$ distances in $\Re^d$ & {\em open} (Conj.\ref{con:geo}) & $1+\varepsilon$ (Sec. 4.3) & n/a \\
\hline
\hline
Online & locally bounded weights & n/a & $O(\log \Delta)$ (Sec. 3.1) & $\Omega(\log \Delta)$ (Sec. 3.1)\\
\hline
\end{tabular}
}
}
\caption{\label{tab:overview}
Overview of results for different variants of the freeze-tag problem. ``LB'' indicates a lower bound; for stars, $\rho$ denotes the number of robots at each leaf; $\Delta$ is the maximum degree of the graph.
}
\end{table}


\subsection{Preliminaries}
Let $R=\{v_0,v_1,\ldots,v_{n-1}\}\subset {\cal D}$ be a set of $n$
{\em robots\/} in a {\em domain\/} ${\cal D}$.
We assume that the robot at $v_0$ is the {\em source robot\/},
which is initially awake; all other robots are initially asleep.
Unless stated otherwise (in Section~\ref{subsec:online}), 
we consider the offline 
version of the problem, in which each
awake robot is aware of the position of all other robots, and
is able to coordinate its moves with those of all the others.
We let $d(u,v)$ indicate the distance between two points,
$u,v\in {\cal D}$.
We study two cases,
depending on the nature of the domain ${\cal D}$:
\begin{specialitemize}
\item
  The space ${\cal D}$ is specified by a graph $G=(V,E)$, with
  nonnegative edge weights.  The robots $v_i$
  correspond to a subset of the vertices, $R\subseteq V$, possibly with
several robots at a single node.
  In the special case in which $G$ is a star,
  the induced metric is a {\em centroid metric\/}.
\item
  The space ${\cal D}$ is a $d$-dimensional geometric space with
  distances measured according to an $L_p$ metric.  We concentrate
  on Euclidean spaces, but our results apply more generally.
\end{specialitemize}
A solution to the FTP can be described by a {\em
  wake-up tree\/} ${\cal T}$ which is a directed binary tree, rooted at
$v_0$, spanning all robots $R$. For any robot $r$, its {\em wake-up path}
is the unique path in this tree that connects $v_0$ to $r$.
If a robot $r$ is awakened by robot
$r'$, then the two children of robot $r$ in this tree are the robots
awakened next by $r$ and $r'$, respectively.
Our objective is to determine an optimal wake-up tree, ${\cal T}^*$,
that minimizes the {\em depth\/}, that is the length of the longest (directed)
path from $v_0$ to a leaf (point of $R$).  We also refer to the depth
as the {\em makespan\/} of a wake-up tree.  We let $t^*$ denote the
optimal makespan (the depth of ${\cal T}^*$).  Thus, the FTP
can also be succinctly stated as a graph optimization problem:
In a complete weighted graph $G$ (the vertices correspond to robots and
edge weights represent distances between robots), find a binary
spanning tree of minimum depth that is rooted at a given vertex $v_0$
(the initially awake robot).

We say that a wake-up strategy is {\em rational\/} if
(1) each awake robot claims and
travels to an asleep unclaimed robot
(if one exists) at the moment that the robot awakens;
(2) a robot performs no extraneous movement, that is,
if no asleep unclaimed robot exists, an awakened robot without
a target does not move.


The following proposition enables us to concentrate on rational strategies:

\begin{proposition}
\label{prop:nonlazy}
Any solution to the FTP can be transformed into a
rational solution
without increasing the makespan.
\end{proposition}

We conclude the introduction by noting that one readily
obtains an $O(\log n)$-approximation for the FTP.

\begin{proposition}
\label{fact:log-approx}
Any rational strategy for the FTP
achieves an approximation ratio of $O(\log n)$.
\end{proposition}

\begin{proof}
We divide the execution into phases.  Phase 1 begins at time 0 and
ends when the original robot first awakens another robot.  At the end
of Phase~1 there are two awake robots.  Let $n_i$ denote the total
number of robots awake at the end of Phase~$i$.  Phase $i$, for
$i=2,3,\ldots$, begins at the moment Phase~$i-1$ ends, when there are
$n_{i-1}$ awake robots, and the phase
ends at the first moment that each of
these $n_{i-1}$ robots has awakened another robot (i.e., at the
instant when the last of these $n_{i-1}$ robots reaches its target).
Thus, with each phase the number of awake robots at least doubles ($n_i\geq
2n_{i-1}$), except possibly the last phase.  Thus, there are at most
$\lceil \log_2 n\rceil$ phases.  The maximum distance traveled by any
robot during a phase is $diam({\cal D})$.  The claim follows by noting
that a lower bound on the optimal makespan, $t^*$, is given by
$diam({\cal D})/2$ (or, in fact, by the maximum distance from
the source $v_0$ to any other point of~${\cal D}$).
\end{proof}

\section{Star Graphs}
\label{sec:star-graphs}

We consider the FTP on {\em weighted stars\/}, also called {\em
  centroid metrics\/}.  In the general case, the lengths of the spokes
and the number of robots at the end of the spokes vary.

We begin with the simplest case, in which all edges of the star have
the same length, and the awake robot is at the central node $v_0$.  We
start by showing that the natural greedy algorithm is optimal.  The
main idea is to awaken the robots in the most populous leaf.  In any
rational strategy, however, all awake robots return to the root {\em
  simultaneously\/}.  Thus, the optimal algorithm has to break ties:
Assume that the robots are indexed by positive integer numbers.  The
robot with smallest index claims a leaf with the most robots, the
robot with second smallest index claims a still unclaimed leaf with
the most robots, and so forth.  Then all robots travel to their
targeted leaf.

We obtain the following lemma:

\begin{lemma}
\label{lem:greedy-opt-star}
The greedy algorithm for awakening all the robots
in a star with all edges of the same length is optimal.
\end{lemma}

\begin{proof}
  The proof is by an exchange argument.  By
  Proposition~\ref{prop:nonlazy} we consider rational optimal
  strategies.  At each stage of the algorithm, a robot always chooses
  to awaken a branch with the most robots at the end.

  Suppose for the sake of contradiction that we have the particular
  optimal schedule whose prefix is greedy for the longest amount of
  time.  Consider the first step when the algorithm is not greedy.
  That is, a robot chooses branch $e_1$, when there is a branch $e_2$
  with more robots.  Instead, we could swap $e_1$ and $e_2$.  Now the
  robot awakens branch $e_2$, but the ``extra'' robots remain idle
  until the time that branch $e_2$ would be awakened; then a robot
  awakens branch $e_1$ and the extra robots idle on branch $e_2$ are
  activated.  Thus, we have a new optimal solution with an even longer
  greedy prefix, and thus we have shown a contradiction.
\end{proof}

The rest of Section~\ref{sec:star-graphs}
considers stars in which edge lengths vary.
Varying  edge lengths already make the problem NP-hard,
even if the same number of sleeping
robots are located at each leaf.  The FTP on stars
nicely illustrates an important
distinction between the FTP and broadcasting
problems~\cite{focs94*202}, which can be solved to optimality
in polynomial
time for the (more complicated) case of trees.

\subsection{Star Graphs with the Same Number of Robots
on Each Leaf}
\label{subsec:star-graphs-same-number}

In Sections~\ref{subsec:greedy},~\ref{subsec:npc},~and~\ref{subsec:ptas}
we assume that an {\em equal number\/}
$q$ of robots are at each leaf node
of a star graph (centroid metric).
The general case is discussed in
Section~\ref{subsec:14}.

We say that a robot has {\em visited an edge and its leaf,\/}
if it has been sleeping there or traveled there.  We
use the following observation, which follows from another simple
exchange argument.

\begin{lemma}
\label{lem:non-decr}
For any instance of the FTP on stars,
where there is an equal number of robots at each leaf vertex,
there exists an optimal solution such that the lengths of the edges
along any root-to-leaf path in the awakening tree
are nondecreasing.
\end{lemma}

\begin{proof}
The proof is by an exchange argument.
We consider rational strategies.
Suppose for the sake of contradiction that no optimal solution has our
desired property
(i.e., that all root-to-leaf paths in the awakening tree have
edges lengths that are nondecreasing over time).
Consider the
optimal solution, such that this nondecreasing property is obeyed for the
longest possible amount of time, and consider the first edge $e_1$ in
the awakening tree disobeying this property.  That is, a descendant edge
$e_2$ of edge $e_1$
in the awakening tree is shorter than edge $e_1$.
(We can modify edge lengths by a vanishingly small
amount so that there are no ties.)

Instead, we could swap $e_1$ and $e_2$ so that the $q$ robots awakened after
branch $e_2$ do the job of the $q$ robots awakened after branch $e_1$ and
vice versa.  Consider all nodes in the original awakening tree that are
descendants of branch $e_1$ but not $e_2$ after the swap, these nodes
are reached earlier.  Now consider all nodes that are descendants of both
branches $e_1$ and $e_2$; these nodes are reached at the same time before
the and after the swap.  Therefore, we have transformed
this optimal solution into another
optimal solution
whose root-to-leaf paths have nondecreasing edge lengths for an
even  longer  amount of time.
Thus, we obtain a contradiction.
\end{proof}

\subsection{Performance of the Greedy Algorithm
Shortest-Edge-First}
\label{subsec:greedy}

Now we analyze the natural greedy algorithm
Shortest-Edge-First
(SEF).
When an (awake) robot arrives at the
root $v_0$, it chooses the
{\em shortest\/} (unawakened and unclaimed) edge to
awaken next.  Interestingly, this natural greedy algorithm is {\em not\/}
optimal.

The simplest example showing that SEF is
suboptimal is a star with $4$ branches $b_1, \ldots, b_4$
of lengths $1, 1, 1$, and $100$, where one asleep robot is at each leaf.  The
optimal solution has makespan $102$: the first robot awakens branch
$b_1$ then $b_4$, while the robot in $b_1$ awakens $b_2$ and then
$b_3$. On the other hand, the greedy algorithm has makespan $104$: first
$b_1$ is awakened, then $b_2$ and $b_3$ at the same time,
and finally $b_4$.

\begin{figure}[htb]
\centerline{\psfig{figure=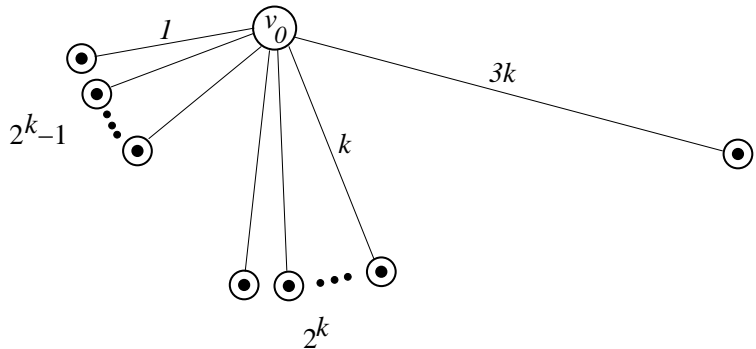,width=.6 \textwidth}}
\caption{Example demonstrating that  Shortest-Edge-First (SEF)
is at best a
$7/3$-approximation.}
\label{fig:shortest-edge-first-stars}
\end{figure}

More generally, we have the following lemma
(Figure~\ref{fig:shortest-edge-first-stars}):

\begin{lemma}
There is a lower bound of $7/3$ on the worst-case approximation factor
of the greedy algorithm.
\end{lemma}

\begin{proof}
Consider a $2^{k+1}$-edge example with
one asleep robot at each leaf. There are $2^k-1$ edges of length $1$,
$2^k$ edges of length $k$, and one edge of length $3k$.  The greedy
algorithm first awakens all robots at short edges.
Thus, at time $2k$, exactly
$2^k$ robots meet at the root, and then they each
go to a (different) edge of length
$k$. Then, at time $3k$, one robot has to travel back to the root and
is sent down to the last sleeping robot at the edge of length $3k$,
where it arrives at time~$7k$.

On the other hand, an optimal solution completes
no later than at time $3k+4$:
this can be achieved by having one robot travel
down the longest edge at time $2$, and another robot travel
down an edge of length $k$ at time $4$, effectively awakening all
short edges while those two long edges are traversed. At time $2k+4$
all short edges and one edge of length $k$ have been awakened and
$2^k$ robots have traveled back to the root (one robot is still
traveling down the edge of length $3k$ and then arrives at time $3k+2$).
We can thus use $2^k-1$ of them to awaken the remaining edges of
length $k$ by the time $3k+4$. Therefore, for large $k$ the ratio of
the greedy solution and the optimal solution tends to $7/3$.
\end{proof}

As it turns out, this example is the
worst case for the greedy algorithm.

\begin{theorem}\label{theorem:7/3}
  For the FTP on stars with the same number of robots
  at each leaf, the performance guarantee of the greedy
  algorithm is $7/3$, and this bound is tight.
\end{theorem}

In order to prove
Theorem~\ref{theorem:7/3},
we first show the following theorem,
which is of independent interest. We define the
{\em completion time of a robot\/}
to be the earliest time when the robot is awake and
resting thereafter, i.e., no longer in motion.
Note that because our strategies are rational,
once a robot rests it never moves again.

\begin{theorem}
 \label{th:average-completion}
  Consider the greedy algorithm Shortest-Edge-First
  on a star for which all leaves have the same number of robots;
  Shortest-Edge-First minimizes the average completion time of all robots.
\end{theorem}

\begin{proof}
  The proof is based on an exchange argument.
  Consider an arbitrary
  solution minimizing the average completion time. Assume that at
  some point in time a robot enters an edge $e_1$ that has a length
  larger than that of a shortest available edge $e_2$. Therefore,
  $e_2$ is chosen at a later point in time.
  There are three cases:

\begin{specialitemize}

\item
{\em Case 1:\/}
  In the tree corresponding
  to the solution, $e_2$ lies in the subtree below $e_1$.
  An exchange of the two edges decreases the average completion time,
  contradicting the optimality of the solution under
  consideration. (This exchange is feasible because both edges have
the same number of robots at their ends.)

\end{specialitemize}

  For $i=1,2$, let $n_i$ denote the number of edges in the subtree
  $T_i$ that is formed by $e_i$ and its descendants.

\begin{specialitemize}

\item
{\em Case 2:\/}
Subtree $T_1$ is larger than $T_2$, that is, $n_1 > n_2$.  Then
exchanging the two edges does not increase the average completion
time.

\item
{\em Case 3:\/}
Subtree $T_1$ is smaller than or equal to $T_2$, that is,
$n_1 \leq n_2$.  Then exchanging the two
subtrees $T_1$ and $T_2$ within the whole tree does not increase the
average completion time.
\end{specialitemize}

By iterating this exchange argument for all three cases, we
finally arrive at the greedy solution, which is therefore
optimal with respect to the average completion time.
\end{proof}

Theorem~\ref{th:average-completion} allows us to proceed.

\begin{proof}[Theorem~\ref{theorem:7/3}]
  Let $m$ be the number of edges in the star and let $q$ be
  the number of sleeping robots
  at each leaf.  In particular, an instance contains $n=1+m\cdot q$
  robots.

Because the {\em average completion time\/} is always a lower bound
  on the {\em maximum completion time\/}, it follows from
  Theorem~\ref{th:average-completion} that the average completion time
  $\bar{C}$ of the greedy solution is a lower bound on the optimal
  makespan.

  We consider rational optimal strategies.
Because a robot only moves if it will later awaken another robot,
all robots  terminate at leaves.
  Moreover, we can assume that {\em for all but one leaf\/}, either all
  $q+1$ robots leave the leaf after awakening, or they all
  stay put. A simple exchange argument proves this observation.

Thus, some leaves have $q+ 1 $ robots ending there,
some leaves have {\em no\/} robots ending there,
and a single leaf may have {\em some\/} robots that end there
and  {\em some\/} that travel to other leaves.
In particular, let
$p:=\lfloor n/(q+1)\rfloor  $
be the number of leaves with
$q+ 1 $
robots ending at them.
Exactly $p (q+ 1) $
robots end at these leaves.  Therefore, the remaining
$n - p (q+ 1) < q+ 1 $
robots end at the single leaf for which (possibly)
some but not all robots depart  to awaken other branches.

We assume that the edges $e_i$, $i=1,2,\ldots,m$ are indexed in
order of nondecreasing lengths $\ell (e_i)$.
The Shortest-Edge-First strategy therefore awakens
the edges in this order.
Because $p$ leaves have $q+ 1 $
robots ending at them (the leaves on edges
$e_{m-p+1}, e_{m-p+2}, \ldots, e_{m} $),
the last robot to awaken anyone
is one of the robots from leaf at $e_{m - p}$.
This last robot $r $
awakens edge $e_m $,
at which point the algorithm terminates.

Let $T$ denote the time when robot $r$
departs from the leaf of edge
$e_{m-p}$
in order to travel back to the root and then to the end of
edge $e_m$. The makespan of the greedy solution is
\begin{equation}\label{eq:upper-bound}
T + \ell (e_{m-p}) +\ell (e_m)\enspace.
  \end{equation}
  By construction, $T$ is a lower bound on the completion time of each
  robot in the greedy solution because
no robot rests until after time $T $.
Thus, we obtain the following lower bounds on the makespan $t ^ * $:
$$
T \leq \bar{C} \leq t^*
\mbox{ and }
\ell (e_m) \leq t^*.$$

It remains to be shown that
 $
\ell(e_{m-p}) \leq t^*/3.
 $
At the end of the {\em optimal\/} solution there are $p$ leaves with
  $q+1$ robots. Therefore, there must be an edge $e_i$ with
  $i\in\{m-p,m-p+1,\ldots,m\}$ and less than $q+1$ robots at its end,
  because $|\{m-p,m-p+1,\ldots,m\}|=p+1$.  Without loss of generality,
  one robot must have traveled down this edge in order to unfreeze the
  $q$ robots at its end and then traveled back to the root in order to
  travel down another edge $e_j$.  Lemma~\ref{lem:non-decr} yields
  $\ell (e_j) \geq \ell (e_i)$, so that the value of any solution is at least
  \[
    2\cdot \ell (e_i)+\ell (e_j) \geq 3\cdot \ell (e_{m-p})\enspace.
  \]
Thus, $t^*\geq 3\cdot \ell(e_{m-p})$,
implying that the makespan of the greedy solution is at most
$t^*+(t^*/3)+t^*=7t^*/3$, completing the proof.
\end{proof}

We conclude our discussion of SEF
by noting a result that will be needed
for the proof of Theorem~\ref{th:star.ptas}.

\begin{corollary}
\label{cor:greedy}
  The makespan of the greedy solution is at most $t^* + 2\ell_{\max}$
  where $\ell_{\max}:=\max_i \ell (e_i)$.
\end{corollary}

\begin{proof}
  This follows from Equation~\eqref{eq:upper-bound} because $T$ is a lower
  bound on $t^*$.
\end{proof}

\subsection{NP-Hardness}
\label{subsec:npc}
We saw in the previous section that the greedy algorithm may not
find an optimal solution. Here we show that it is unlikely
that any other polynomial algorithm can always find an optimum.

\begin{theorem}
\label{th:star.npc}
  The FTP is strongly NP-hard, even for the
  special case of weighted stars with one (asleep) robot at each leaf.
\end{theorem}

\begin{proof}
  Our reduction is from {\sc Numerical 3-Di\-men\-sio\-nal Matching} (N3DM)~\cite{gj-cigtn-79}:
\smallskip

  \noindent {\bf Instance}: Disjoint sets $W$, $X$ and $Y$, each
containing $n$
  elements, a size $a_i\in Z^+$ for each element $i\in W$, a size $b_j\in
  Z^+$ for each element $j\in X$, a size $c_k\in Z^+$ for each element $k\in
  Y$, such that $\sum_{i\in W}a_i+\sum_{j\in X}b_j+\sum_{k\in Y}c_k=d\,n$
  for a target number $d\in Z^+$.
\smallskip

  \noindent {\bf Question:} Can $W\cup X\cup Y$ be partitioned into $n$
  disjoint sets $S_1,S_2,...,S_n$, such that each $S_h$ contains
  exactly one element from each of $W$, $X$, $Y$ and such that for
  $1\leq h\leq n$, $a_{i_h}+b_{j_h}+c_{k_h} = d$?

\medskip

 See Figure~\ref{fi:star} for the overall idea of the reduction.
  For technical reasons we assume without loss of generality that the
  size of each element from $W\cup X\cup Y$ is at most $d$.
  Moreover, we can assume without loss of generality
  that $n=2^K$ for some $K\in N$ --- the number of elements in $W$,
  $X$, and $Y$ can be increased to the nearest power of $2$ by adding
  elements of size $d-2$ to $W$ and elements of size $1$ to $X$ and
  $Y$; notice that this does not affect the value ``yes'' or ``no'' of
  the instance.

  Let $\varepsilon$ be a sufficiently small number ($\varepsilon<
  1/(2K)$ suffices), and let $L$ be sufficiently large, e.g.,
  $L:=15d$.  Consider a designated root node with an awake robot, and
  attach the following edges to this root:
  \begin{itemize}
  \item $n-1$ edges of length $\varepsilon$; $E$ denotes the robots at these leaves, along with
the robot at $v_0$.
  \item $n$ edges of length $\alpha_i:=a_i/2-\varepsilon K+d$,
 $i=1,\ldots,n$; $A$ denotes the robots
    at these ``$A$-leaves''.
  \item $n$ edges of length $\overline{\alpha_i}:=L-a_i-2d$, for
    $i=1,\ldots,n$; $\overline{A}$ denotes the robots at
    these ``$\overline{A}$-leaves''.
  \item $n$ edges of length $\beta_j:=b_j/2+2d$, for
    $j=1,\ldots,n$; $B$ denotes the robots at these
    ``$B$-leaves''.
  \item $2n$ edges, two each of length $\gamma_k:=L-7d+c_k$, for
    $k=1,\ldots,n$; $C$ denotes the set of robots at these
    ``$C$-leaves''.
  \end{itemize}

\begin{figure}[thbp]
\begin{center}
\leavevmode
\epsfig{file=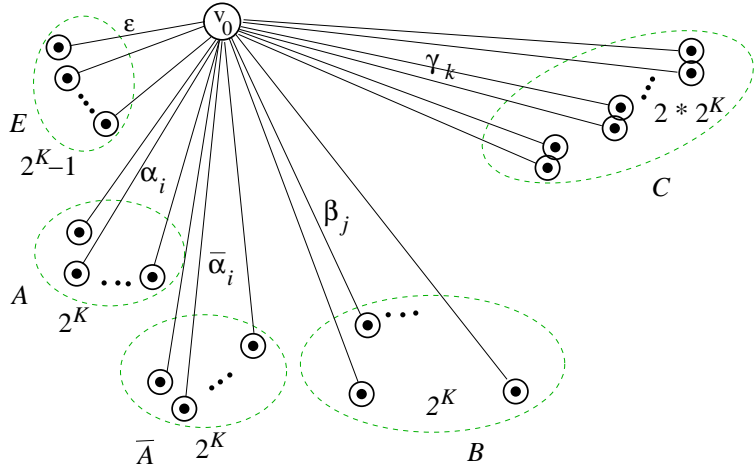,width=0.6\columnwidth}   
\caption{NP-hardness of Freeze-Tag for stars: In any good solution,
a robot awakening one of the robots in set $C$ must have visited
the sets $A$ and $B$ precisely once. This means that there is a cheap
solution for this class of
FTP instances, iff the elements of the sets $A$, $B$, $C$ can be grouped
such that $\alpha_{i_h}+\beta_{j_h}+\gamma_{k_h}=d$.
}
\label{fi:star}
\end{center}
\end{figure}

  We claim that there is a schedule to awaken all robots within time
  $L$, if and only if there is a feasible solution to the N3DM
  instance.

  It is straightforward to see that the ``if'' part
  holds: Let $S_1,S_2,...,S_n$ be a feasible solution to the N3DM
  instance.  Using a binary
  tree on the set $E$ (a ``greedy cascade on $E$''),
  we can bring all $n$ robots in $E$ to the root at time
  $2\varepsilon K=2\varepsilon \log n$.
  These $n$ robots are sent to the $A$-leaves. Now there
  are $2n$ robots available, two each will get back to the root at
  time $a_i+2d$, $i=1,\ldots,n$. One of each pair is sent down the
  edge of length $\overline{\alpha_i}$, so that the whole set
  $\overline{A}$ gets awakened just in time $L$. The remaining $n$
  robots (one for each $a_i$, call this robot $A_i$) get sent to wake
  up the robots of set $B$, such that $A_{i_h}$ is assigned to an edge
  of length $\beta_{j_h}$, if $a_{i_h}$ and $b_{j_h}$ belong to the
  same set $S_h$.  This gets two robots for each $h$ (say,
  $A_{i_h}^{(1)}$ and $A_{i_h}^{(2)}$) back to the root at time
  $a_{i_h}+b_{j_h}+6d$. Send those two robots down the two edges of
  length $\gamma_{k_h}$.  Because $a_{i_h}+b_{j_h}+c_{k_h}=d$, all
  robots in $C$ are awake at time $L$.

  To see that a feasible schedule implies a feasible solution of the N3DM
  instance,
  first observe that no robot in $F=\overline{A}\cup C$ can
  wake up any other robot, as the corresponding edges are longer than
  $L/2$.  Moreover, the same argument implies that no two robots in
  $F$ can be awakened by the same robot.  Because the total number of
  robots is precisely $2|F|=6n$, we conclude that each robot in $E\cup
  A\cup B$ must wake up a different robot in $F$.

  Clearly, no robot in $B$ can wake up a robot in $\overline{A}$ by
  the deadline $L$.  Thus, the robots in $\overline{A}$ are awakened
  by a set of $n$ robots $\tilde{A}\subset E\cup A$.
  Notice that a robot in $\tilde{A}$ can neither visit a
  $B$-leaf nor two $A$-leaves and
  still meet the deadline.

  The $2n$ robots in $C$ must be awakened by the $2n$ robots in
  $B\cup(E\cup A)\setminus\tilde{A}$.  Because none of them has enough
  time to visit two $B$-leaves, each must visit exactly one $B$-leaf
  and then, by Lemma~\ref{lem:non-decr}, travel immediately to a
  $C$-leaf.  We can assume without loss of generality (by a simple
  exchange argument) that each pair of robots that has visited the
  same $B$-leaf is assigned to a pair of $C$-leaves at the same
  distance.

  As explained above, each robot in $\tilde{A}$ can visit at most one
  $A$-leaf; the same is true for all robots in $(E\cup
  A)\setminus\tilde{A}$, because each must visit one $B$-leaf and one
  $C$-leaf afterwards.  Because there are $2n$ robots in $E\cup A$ and
  also $2n$ visits to $A$-leafs, each robot in $E\cup A$ must visit
  exactly one $A$-leaf.

  We next argue that, without loss of generality, a feasible solution
  uses a greedy cascade in the beginning to bring all $n$ robots in
  $E$ to the root at time $2\varepsilon K$. As described above, the
  greedy cascade guarantees that, later, each pair of robots
  returning from an $A$-leaf at distance $\alpha_i$ arrives at the
  center node at time $a_i+2d$. On the other hand, such a pair cannot
  arrive before time $2\alpha_i=a_i+2d-2\varepsilon K > a_i+2d-1$.
  Because the deadline $L$ as well as all remaining travel times to
  $\overline{A}$-leaves or $B$ and $C$-leaves are integral, the claim
  follows.

  A simple exchange argument yields that, without loss of generality,
  the robots in $\overline{A}$ are awakened by the robots in $A$,
  i.\,e., $\tilde{A}=A$. Thus, each robot in $E$ travels to one
  $A$-leaf, then to a $B$-leaf and finally to a $C$-leaf.  The time at
  which a robot in $E$ who has visited edges of length $\alpha_i$ and
  $\beta_j$ arrives at a $C$-leaf at distance $\gamma_k$ is
  $L-d+a_i+b_j+c_k$. Therefore, a schedule that awakens all robots by
  time $L$ implies a partition $S_1,\dots,S_n$ with
  $a_{i_h}+b_{j_h}+c_{k_h}=d$ for all $h$.
\end{proof}

As the problem {\sc 3-Partition} is strongly NP-complete, it is straightforward
to convert the weighted stars in the construction into unweighted trees by
replacing weighted edge by an unweighted path.

\begin{corollary}
\label{cor:unw.trees}
The FTP is NP-hard, even for the
  special case of unweighted trees with one (asleep) robot at each leaf.
\end{corollary}

\subsection{PTAS}
\label{subsec:ptas}
We give a PTAS for the FTP on
weighted stars with one awake robot at the central node, $v_0$, and an equal
number $q$ of sleeping robots at each leaf. The underlying basic idea
is to partition the set of edges into ``short'' and ``long'' edges. The
lengths of the long edges are rounded such that only a constant number
of different lengths remains. The approximate positions of the long
edges in an optimal solution can then be determined by complete
enumeration. Finally, the short edges are ``filled in'' by a variant
of the greedy algorithm discussed in Subsection~\ref{subsec:greedy}.
During each
step we may lose a factor of $1+O(\varepsilon)$, such that the
resulting algorithm is a $(1+O(\varepsilon))$-approximation algorithm,
so we get a polynomial-time approximation scheme.

Similar techniques have been applied for other classes of problems
before, e.g., in the construction of approximation schemes for
machine scheduling problems (see for example~\cite{hs87-udaasp}).
However, the new
challenge for the problem at hand is to cope with the
awakened robots at short edges whose
number can increase geometrically over time.

Let $T\leq t^*$ be a lower bound on the makespan, $t^*$, of an optimal solution.
For our purpose, we can set $T$ to $3/7$ times the makespan of the
greedy solution, which can be determined in polynomial time.
For a fixed constant $\varepsilon>0$, we partition the set of edges
$E$ into two subsets
\begin{eqnarray*}
  S&:=&\{e\in E\mid \ell (e)\leq \varepsilon T\}\qquad\text{and}\qquad\\
  L&:=&\{e\in E\mid \ell (e) > \varepsilon T\}\enspace.
\end{eqnarray*}
We call the edges in $S$ \emph{short} and the edges in $L$
\emph{long}.
We modify the given instance by rounding up the length
of each long edge to the nearest multiple of $\varepsilon^2 T$.

\begin{lemma}
\label{lem:rounding}
  The optimal makespan of the rounded instance is at most
  $(1+O(\varepsilon))t^*$.
\end{lemma}

\begin{proof}
Consider the awakening tree corresponding to an optimal solution of the
  original instance.  On any root-to-leaf path in the awakening tree,
there can be at most $O(1/\varepsilon)$ long edges.
(This is because
$T\leq t^* \leq \frac{7}{3} T$,
and long edges have length at least
$\varepsilon T$.)
In the rounding step we increase the length of a long edge by at most
  $\varepsilon^2 T$. Therefore the length of any path, and thus the
  completion time of any robot in the solution given by the tree, is
  increased by at most $O(\varepsilon)\cdot T$. Because $T$
is bounded by
$T\leq t^* \leq \frac{7}{3} T$, the claim follows.
\end{proof}

Any solution to the rounded instance with makespan~$t$ induces a solution
to the original instance with makespan at most~$t$.  Therefore, it
suffices to
construct a $(1+O(\varepsilon))$-approximate solution to the rounded
instance.  In the following we only work on the rounded
instance and refer to it as \emph{instance~$I$}.

\begin{lemma}\label{lem:rounded-start}
  There exists a $(1+\varepsilon^2)$-approximate solution (which is
  possibly not a rational strategy) to instance~$I$ that meets the
  requirement of Lemma~\ref{lem:non-decr}, such that for each long
  edge the point in time that it is entered by a robot (its ``start
  time'') is a multiple of $\varepsilon^2 T$.
\end{lemma}

\begin{proof}
  An optimal solution to instance~$I$ can be modified to meet the
  requirement of the lemma as follows.  Because the schedule obeys the
  structure from Lemma~\ref{lem:non-decr}, any root-to-leaf path in
  the awakening tree first visits all short edges before all long
  edges.  Whenever a robot wants to enter a long edge, it has to wait
  until the next multiple of $\varepsilon^2 T$ in time.  Because the
  lengths of long edges are multiples of $\varepsilon^2 T$ all
  subsequent long edges are entered at times that are multiples of
  $\varepsilon^2 T$.  Therefore this modification increases the
  makespan of the solution by at most $\varepsilon^2 T$.
\end{proof}

In the remainder of the proof we consider an optimal solution to
instance~$I$ meeting the requirements of Lemma~\ref{lem:non-decr} and
Lemma~\ref{lem:rounded-start}.  The makespan of this solution is
denoted by $t^*$.  Notice that both the number of different lengths of
long edges and the number of possible start times are in
$O(1/\varepsilon^2)$. The positions of long edges in an optimal
solution can be described by specifying for each possible start time
and each edge length the number of edges of this length that are
started at that time. Because each such number is bounded by the total
number of edges $n$, there are at most $n^{O(1/\varepsilon^4)}$
possibilities, which can be enumerated in polynomial time.

This enumerative argument allows us to assume that we have guessed the
correct positions of all long edges in an optimal solution. Again, 
we can assume by Lemma~\ref{lem:non-decr} that any
root-to-leaf path in the awakening tree first visits all short edges
before all long edges.  Therefore, the long edges are grouped together
in {\em subtrees.\/} We must fill in the short edges near the root to
connect the root of the awakening trees to the subtrees consisting of
long edges.  Given the start times $S(e)$ of the long edges $e\in L$,
we group them into subtrees as follows. One by one, in order of
nondecreasing start times, we consider the long edges.  If an edge $e$
has not been assigned to a subtree yet, we declare it the root of a
new subtree. If there are long edges $e'$ with $S(e')=S(e)+2\ell (e)$
that have not been assigned to a subtree yet, we assign at most $q+1$
of them as children to the edge $e$.

Let $p$ be the number of resulting subtrees and denote the start times
of the root edges by $S_1,\ldots,S_p$, indexed in nondecreasing order.
Notice that, although the partition of long edges into subtrees is in
general not uniquely determined by the vector of start times
$(S(e))_{e\in L}$, the number of subtrees $p$ and the start times
$S_1,\ldots,S_p$ are uniquely determined.

It remains to fill in all short edges. This can be done by
the following variant of the greedy algorithm: We set
$S_{p+1}:=\infty$ and $i:=1$ in the beginning. Assume that a robot,
coming from a short edge, gets to the central node at time $t$.
\begin{itemize}
\item If $t\geq S_i+2\varepsilon T$, then send the robot into the
  $i$th subtree and set $i:=i+1$.
\item Else, if there are still short edges to be visited, then send
  the robot down the shortest of those edges.
\item Else, if $i\leq p$, then send the robot into the $i$th
  subtree and set $i:=i+1$.
\item Else stop.
\end{itemize}

\begin{lemma}\label{lem:greedy-makespan}
  The above generalized greedy procedure yields a feasible solution to
  instance~$I$ whose makespan is at most $t^* +4\varepsilon T\leq
  (1+4\varepsilon)t^*$.
\end{lemma}

\begin{proof}
  We first argue that the solution computed by the generalized greedy
  procedure is feasible, i.\,e., all robots are awakened.  We thus
  have to show the following: When a robot is sent into the $i$th
  subtree then either 
  all short edges have been visited or
  there is at least one other robot traveling along a short edge
  (which will take care of the remaining subtrees and/or short edges).
  
  Assume by contradiction that this condition is violated for some
  $i\in\{1,\dots,p\}$.  Let $t$ denote the point in time when the
  generalized greedy procedure sends the last robot traveling on short
  edges into the $i$th subtree.  We consider a (partial) solution to
  a modified instance~$I'$ which is obtained by replacing the first
  $i-1$ subtrees in the solution computed by the generalized greedy
  procedure until time $t$ by subtrees consisting of new short edges.
  These new edges and subtrees are chosen such that the resulting
  solution until time $t$ has the Shortest-Edge-First property, i.e.,
  it is a (partial) greedy solution.
  
  To be more precise, the construction of the modified instance~$I'$
  and solution can be done as follows: Whenever the generalized greedy
  procedure sends a robot into one of the first $i-1$ subtrees, we
  replace the first edge of this subtree in the solution to the
  modified instance~$I'$ by a new edge whose length equals the length
  of the shortest currently available edge.  Moreover, whenever a
  robot belonging to the modified subtree arrives at the center node
  before time $t$, we add a new short edge to the modified
  instance~$I'$ whose length equals the length of the shortest
  currently available edge and assign the robot to it.
  
  Let $k$ be the number of awake robots in this greedy solution for
  the modified instance~$I'$ at time $t$.  Notice that all $k$ robots
  belong to one of the modified subtrees.

  The optimal solution to instance~$I$ until time
  $t':=t-2\varepsilon T$ induces a solution $\sigma$ to the modified
  instance~$I'$ until time $t'$ by again replacing the first $i-1$ subtrees
  of long edges by the corresponding subtrees of short edges.

  We claim that there are at least $k+1$ awake robots in $\sigma$ at
  time $t'$.  Notice that the first $i-1$ subtrees are started at
  least $2\varepsilon T$ time units earlier than in the greedy
  solution (by construction of the generalized greedy procedure).
  Thus, the number of awake robots in these subtrees at time $t'$ is
  at least $k$.  Moreover, because the optimal solution awakens all
  robots, there must be at least one additional awake robot at time
  $t'$ (taking care of the remaining edges).

  However, in order to get $k+1$ awake robots, $\lceil k/q\rceil$
  leaves must have been visited.  Consider the $\lceil k/q\rceil$
  shortest edges of the modified instance~$I'$.  It follows from the
  discussion in the last paragraph that it takes at most $t'$ time
  units to visit all of these edges.  On the other hand, the makespan
  of the greedy solution is larger than $t$ because the number of awake
  robots in the greedy solution at time $t$ is only $k$.  Because we
  only consider short edges of length at most $\varepsilon T$ and
  because $t-t'=2\varepsilon T$, this is a contradiction to
  Corollary~\ref{cor:greedy}.

  So far we have shown that the solution computed by the generalized
  greedy procedure visits all leaves.  It remains to show that its
  makespan is at most $t^* +4\varepsilon T$.

  Notice that the length of the time interval between two visits of a
  robot traveling on short edges to the central node is at most
  $2\varepsilon T$.  Therefore, a robot is sent into the $i$th
  subtree before time $S_i+4\varepsilon T$, for $i=1,\dots,p$.  As a
  consequence, the robots at each long edge are awake before time
  $t^*+4\varepsilon T$.

  Finally, the same argument as in the feasibility proof above shows
  that all robots at short edges are awake at time
  $t:=t^*+2\varepsilon T$.  This completes the proof.
\end{proof}

We summarize:

\begin{theorem}
\label{th:star.ptas}
  There exists a polynomial-time approximation scheme for the
  FTP on (weighted) stars with the same number of robots at each leaf.
\end{theorem}

\subsection{Any Number of Robots at Each Leaf}
\label{subsec:14}

We give a constant-factor approximation algorithm for the FTP on general stars
(centroid metric), where edge lengths may vary and
leaves may have different numbers of
asleep robots.  Interestingly,
we obtain this $O(1)$-approximation algorithm by merging (interleaving) two
natural algorithms, each of which may perform poorly:

\begin{specialitemize}

\item
The {\em Shortest Edge First (SEF)\/}
strategy, where an awake robot
at $v_0$ considers the set of shortest edges leading to
asleep robots, and chooses one with a maximum number of asleep robots.

\item
The {\em Repeated Doubling (RD)\/}
strategy, where the edge
lengths traversed repeatedly (roughly) double in size, and
in each length class, edges are selected by decreasing number
of asleep robots.  (A formal definition of the RD strategy is
given later.)
\end{specialitemize}

\noindent
Each of these algorithms is only a $\Theta(\log n)$-approxi\-ma\-tion, but
their {\em combination\/} leads to an $O(1)$-approxi\-ma\-tion.

We first consider the Shortest-Edge-First strategy.

\begin{lemma}
The SEF strategy leads to a $\Theta(\log n)$-approximation.
\end{lemma}

\begin{figure}[tbh]
\centerline{\psfig{figure=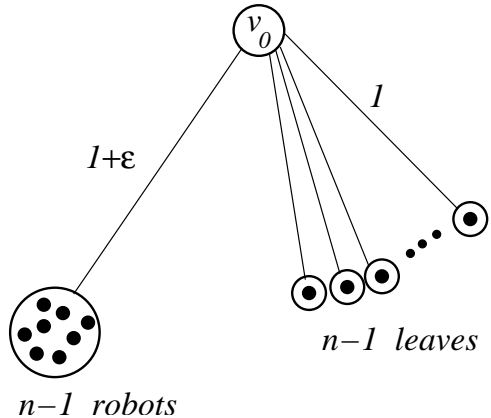,scale=0.5}}
\caption{Example in which Shortest-Edge-First yields a solution of
makespan $\Theta( \log n )$, while the optimal is $O(1)$.}
\label{fig:shortest-edge-first}
\end{figure}

\begin{proof}
Consider a tree having one edge of length $1+\varepsilon$ that leads to a leaf with $n-1$ robots,
and $n-1$ edges, each of length $1$, with one robot at each of the
leaves; see Figure~\ref{fig:shortest-edge-first}.
The Shortest-Edge-First strategy has makespan
$\Theta(\log n)$, whereas an optimal strategy has
makespan $O(1)$.  Thus, SEF is an
$\Omega(\log n)$
approximation algorithm.
By Proposition~\ref{fact:log-approx}, any rational strategy is an
$O({\log n})$
approximation algorithm.
\end{proof}

\begin{figure}[htb]
\centerline{\psfig{figure=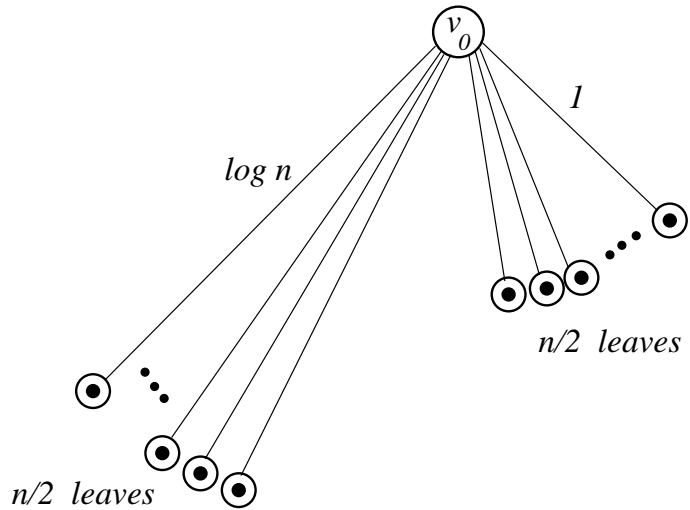,scale=0.5}}
\caption{Example in which Repeated Doubling
yields a solution of $\Theta(\log^2 n)$,
while the optimum is $O (\log n)$.}
\label{fig:repeated-doubling-bad}
\end{figure}

Next we consider the Repeated Doubling strategy.  A robot at $v_0$
faces the following dilemma: should the robot choose a short edge
leading to a small number of robots (which can be awakened quickly) or
a long edge leading to many robots (but which takes longer to awaken)?
There are examples that justify both decisions, and where a wrong
decision can be catastrophic.
See Figure~\ref{fig:repeated-doubling-bad}.

We begin our analysis by assuming that all branches have lengths that
are powers of $2$.  This assumption is justified because we can take
an arbitrary problem and stretch all the edges by at most a factor of
$2$.  Now any optimal solution for the original problem becomes a
solution to this stretched problem, in which the makespan is increased
by at most a factor of $2$.  Thus, a $k$-approximation to the stretched problem
is a $2k$-approximation to the original problem.

Thus, we have reduced the problem on general stars to the problem on
stars whose edge lengths are powers of $2$.  We partition the edges
into \emph{length classes}.  Within each length class, it is clear
which edge is the most desirable to awaken: the one housing the most
robots.  However, how can we choose which length class the robot
should visit first?  Suppose that an optimal algorithm chooses an edge
of length $2^ j $
at some point in time.  We can visit edges of lengths
$1,2,4,8, \ldots, 2^ j $
and only increase the makespan by a factor of $3$.  That is, we
use repeated doubling to ``hedge our bets'' about what is the best path to
take next.  However, if the right choice to make is to awaken robots
in a nearby length class, then we may suffer by sending robots on a
repeated-doubling trajectory to long edges.

In summary, the Repeated Doubling (RD) algorithm is as follows.
When a robot wakes up,
it awakens the most desirable edge in length class
$1,2,4,8, \ldots$.  When the robot
runs out of length classes, it starts the repeated
doubling process anew.

The Repeated Doubling strategy may have poor performance:

\begin{lemma}
The RD strategy yields an $\Theta(\log n)$-approximation.
\end{lemma}

\begin{proof}
Consider a star having $n/2$ edges of length $1$ and
$n/2$ edges of length $\log n$, with a single robot at each leaf.
Refer to Figure~\ref{fig:repeated-doubling-bad}.  
The optimal solution has makespan  $\Theta ( {\log n } )$ ---
first the robots awaken all the short branches
(in time  $\Theta ( {\log n } )$),
and then $n/2$ robots awaken the long branches
(again in time $\Theta (\log n)$). The RD strategy, on the
other hand, has makespan $\Theta (\log^2 n)$.
Thus, the
RD strategy is an
$\Omega (\log n) $
approximation.
By Proposition~\ref{fact:log-approx}, any rational strategy is an
$O ( {\log n} ) $ approximation algorithm,
establishing our bound.
\end{proof}

We now merge these two previous strategies to obtain what we call
the {\em Tag-Team Algorithm\/}:
When a robot is first awakened, it awakens one edge
in each length class
$1,2,4,8, \ldots$.
{\em Before\/} each doubling step, the robot awakens
the shortest edge that it can find.
When the robot runs out of length classes, it starts the repeated
doubling process anew. Naturally,
the robot skips any length class no longer containing edges.

\begin{theorem}
\label{thm:tag-team}
The Tag-Team algorithm gives a 14-approximation for the
FTP on centroid metrics (general stars).
\end{theorem}

\begin{proof}
  We begin by restricting ourselves to the special case in which all
  edge lengths are powers of 2; because any general instance can be
  transformed to this special case, while at most doubling the edge
  lengths, this restriction results in at most doubling the cost of a
  solution.
  
  Consider an optimal solution given by a wake-up tree~${\cal T}^*$.
  We can assume without loss of generality that an edge is awakened
  before all other edges in the same length class with a smaller
  number of robots.  Moreover, if there are several edges with the
  same number of robots in a length class, we break ties and assume
  that the Tag-Team algorithm visits these edges in the same order as
  the optimal solution does.
  
  We show by induction that if in the optimal awakening tree ${\cal
    T}^*$ an edge $e$ is awakened at time $t$, then the Tag-Team
  algorithm awakens this edge $e$ at or before time~$7t$.
  
  Suppose that in the optimal awakening tree ${\cal T}^*$ at time $t$
  a robot ${r}$ awakens the robots ${r_1}, {r_2}, \ldots, {r_k}$ at the
  end of an edge $e$, where $e$ has length $\ell (e)$.  Consider the
  next edge that each of the robots ${r_1}, {r_2}, \ldots, {r_k}$
  awakens in the optimal awakening tree ${\cal T}^*$.  Specifically,
  suppose that in ${\cal T}^*$, robot ${r_i}$ travels to an edge $e_i$
  of length ${\ell (e_i)}$.  That is, in ${\cal T}^*$ at time $t +\ell
  (e) +\ell (e_i)$, robot $r_i$ awakens the robots at the end of edge
  $e_i$.
  
  Now we consider when these robots are awakened in the Tag-Team
  algorithm.  By induction, suppose that robot $r$ was awakened at or
  before time $7t$.  In the tag-team algorithm each of the awakened
  robots ${r_1}, {r_2}, \ldots, {r_k}$ performs a repeated doubling
  trajectory, ultimately awakening the edge in the appropriate length
  class.  The worst case is when the edge taken in the SEF branches
  has the same length as the edge taken in the RD branches.  Thus, two
  edges of length $2 ^ j $ are traversed (for $j =1, 2, \ldots $), one
  during a RD step and one during an SEF step, and each edge is
  traversed in both directions.  Therefore, either $r_i$ awakens $e_i$
  by time
  \[
    7t +\ell (e)+ 2\cdot 2\cdot (1+2+4+\dots+\ell (e_i)/2) +
    2\ell (e_i) +\ell (e_i)
    \, \leq \,  7t +\ell (e)+ 7\ell (e_i) \, ,
  \]
  or edge $e'_i$ was already awakened. That is, even without knowing
  the edge class where $r_i$ should go directly, it gets there
  eventually.  But what about the original robot $r$, which continues
  its repeated doubling trajectory to larger edges when in ${\cal
    T}^*$ robot $r$ visits a smaller edge?  Robot $r$ plays tag-team
  awakening at least one robot on the smallest edge possible, and this
  new robot performs $r$'s duties. Thus, this robot in time $\leq 7t
  +\ell (e)+ 7 \ell (e_i) $ awakens the edge class that $r$ would
  awaken in ${\cal T}^*$.
\end{proof}

At this point it is still open how this approximation factor can be 
improved. In fact, we conjecture that there is a $(1+\varepsilon)$-approximation:

\begin{conj}
\label{con:uneven}
There is a PTAS for the freeze-tag problem on weighted stars with
not necessarily the same number of robots at each leaf.
\end{conj}


\section{General Graphs}

Now we discuss the FTP
on general graphs $G=(V,E)$ with nonnegative edge weights
$\ell (e)$.  We let $\delta(v)$ denote the degree of $v$ in $G$.


\subsection{A Competitive Online Algorithm}
\label{subsec:online}
As Theorem~\ref{th:star.npc} illustrates,
even the presence of a single vertex $v$ with a high degree 
causes the problem to be NP-hard, showing
that the resulting choices may be difficult to resolve. 
This makes it plausible
that the complete absence of high-degree
nodes could make the problem more tractable.
As we will see later on, this is not the case:
Even for graphs of maximum degree 5, finding a solution
within 5/3 of the optimum is NP-hard.

However, it is not hard to see that a sufficient
number, $r(v_i)$, of robots at each vertex $v_i$ yields an easy problem:

\begin{lemma}\label{lem:BFS}
Suppose $r(v_0)\geq \delta(v_0)$ for the source node $v_0$,
and $r(v_i) \geq \delta(v_i)-2$ at any other node $v_i$ in $G$.
Then the FTP can be solved by breadth-first search.
\end{lemma}

This observation is based on the simple fact that any node
in a breadth-first search 
(BFS) tree has minimal possible distance from the root,
making the depth of this tree a general lower bound
on the makespan of a wake-up tree. 
If $r(v_0)\geq \delta(v_0)$ and $r(v_i) \geq \delta(v_i)-2$ for any
$v_i\neq v_0$, we have sufficiently many
robots available to use a BFS tree as the
wake-up tree, and the claim follows.

As we noted in the introduction, the online version of the
FTP is of interest in some potential applications. Using the fact that
BFS uses only local information,
we obtain some simple online results for the FTP, as we now
describe.
We let 
\short{
$\Delta_G:=\max\{\frac{\delta(v_0)}{r(v_0)},\frac{\delta(v_i)-2}{r(v_i)}$, 
$i=1$,$\ldots,n-1\}.$
}
\full{
$$\Delta_G:=\max\{\frac{\delta(v_0)}{r(v_0)},\frac{\delta(v_i)-2}{r(v_i)}, 
i=1,\ldots,n-1\}.$$
}
For an edge-weighted graph $G$ and any node $v$ of $G$, we let
$\rho_v$ denote the maximum ratio of weights for two edges incident on
$v$; i.e., $\rho_v\geq 1$ is the ratio of the maximum edge weight
among edges incident on $v$ to the minimum edge weight among edges
incident on $v$.
We say that $G$ has {\em locally bounded edge weights} if there
exists a constant, $C$, such that $\rho_v\leq C$ for all $v$ in~$G$.

\begin{theorem}\label{th:approx.deg}
Let $G$ be an edge-weighted graph 
with locally bounded edge weights.
Then, there is a linear-time online algorithm
for the FTP on $G$ that guarantees
a competitive ratio of $O(\log \Delta_G)$.
\end{theorem}

\begin{proof}
The idea is to simulate a breadth-first search at
each node: At any vertex $v_i$,
use the robots at $v_i$ to awaken
all robots at neighboring nodes prior to
sending robots to awaken robots at nodes that
neighbor the neighboring nodes of $v_i$.
This is readily achieved
with a binary wakeup tree of 
unweighted depth $O(\log \frac{\delta(v_0)}{r(v_0)})$
for the root $v_0$, and $O(\log \frac{\delta(v_i)-2}{r(v_i)})$
for any other vertex $v_i$, as the vertex used to enter
$v_i$ does not need to be awakened.  Thus, with the assumption
of locally bounded edge weights, the time
needed to do this awakening is 
$O(\log \frac{\delta(v_0)}{r(v_0)})$
(or $O(\log \frac{\delta(v_i)-2}{r(v_i)})$) times the weight of a
minimum-weight edge incident on $v_0$ (or $v_i$).
Thus, each robot is awakened by time $O(\log \Delta_G)$ times
the length of the minimum-weight path from the root to the node
where the robot originally sleeps.
This implies the claim.
\end{proof}

There is an $\Omega (\log \Delta )$ lower bound on the
competitive ratio of any online algorithm, as the following example
shows.
Specifically, $v_0$ has $k$ neighbors, each at distance one
from $v_0$ and each having exactly one sleeping robot.  One of
these neighbors of $v_0$ has adjacent to it a tree with 
diameter $\varepsilon$, having a population of at least $k$ sleeping robots.
An online algorithm has no knowledge of which neighbor of $v_0$ has the
adjacent tree of many sleeping robots; an adversary can make this
neighbor be the last one awakened by the algorithm.  An optimal
offline strategy awakens this neighbor first, then awakens
the neighboring tree of many robots, which then return to
$v_0$ and awaken the rest of $v_0$'s neighbors, in total time $O(1)$.
The online strategy takes time $\Omega(\log \Delta)$.

\subsection{Hardness of Approximation}
As it turns out, there is no realistic hope for a PTAS
on general graphs of bounded degree, even if we go beyond
strictly local, i.e., online procedures:

\begin{theorem}\label{th:no.approx}
It is NP-hard to approximate the FTP
on general weighted graphs within a factor less than 5/3,
even for the case of $\Delta_G=4$ and one robot at each node.
\end{theorem}

\begin{proof}
The reduction is from 3SAT.
Without loss of generality, we assume that we have $n=2^K$
variables. For technical reasons, we
add $n$ clauses of size 2 of the form ``$x$ or not $x$'', one
for each variable.

This instance will be mapped to an
FTP instance on a weighted graph
of bounded degree with one robot per vertex, such
that we have a solution of makespan $3/2+O(\varepsilon\log n)$ if
there is a satisfying truth assignment, and a makespan of at least
5/2 if there is no such truth assignment.
By choosing $\varepsilon=o(\log n)$, this implies that approximating
the resulting class of FTP instances within a factor of less than 5/3
requires finding a satisfying truth assignment, hence the claim.

\begin{figure}[t]
\begin{center}
\leavevmode
\epsfig{file=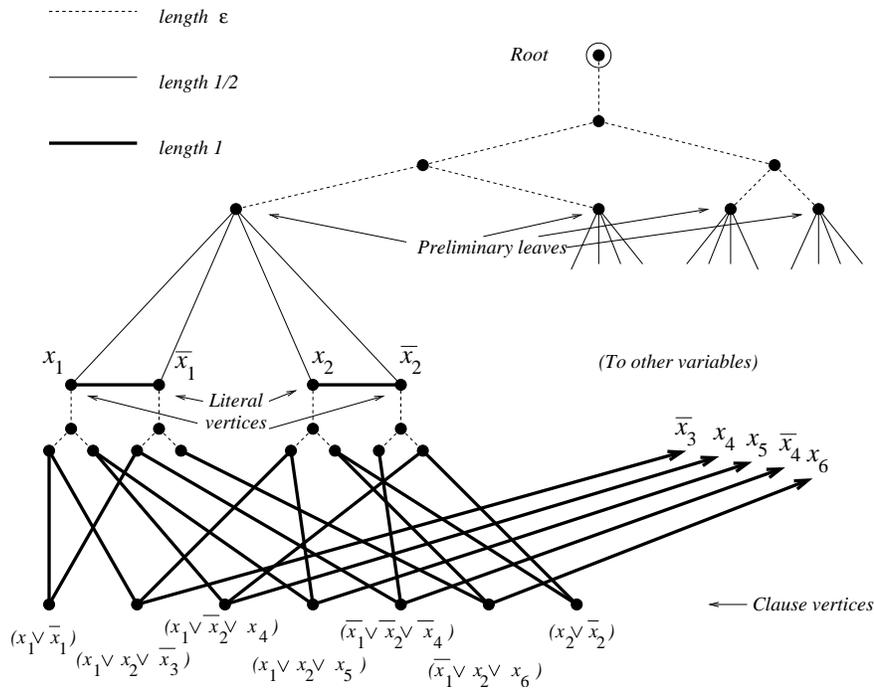,width=.85\columnwidth}   
\caption{NP-hardness of 5/3-approximation of Freeze Tag
in general graphs.}
\label{fig:5.3}
\end{center}
\end{figure}

We now give details of the construction.  See Figure~\ref{fig:5.3}.
{From} the root, build a binary tree of depth $\varepsilon\log(n/2)$,
resulting in two awake robots at each of the $n/2$ {\em preliminary
leaves\/}, after time $\varepsilon\log(n/2)$.

Next group the $n$ variables in an arbitrary way to $n/2$
pairs, and assign a pair to each preliminary leaf.
Each pair of variables is represented by four more vertices,
two corresponding to ``true'', two corresponding to ``false''.
All get connected to the respective preliminary leaf,
using an edge of ``intermediate'' length 1/2.
The two vertices for the same variable get connected to each other,
using an edge of ``long'' length 1.

If $c(x)$ is the number of clauses in which some literal
$x$ occurs, attach a small binary tree of height $O(\varepsilon)$
to allow $c(x)+1$ awake robots at cost $O(\varepsilon)$ when
the literal node is reached. (This does not affect the overall~$\Delta$.)

Finally, add one vertex per clause (including the artificial ones
stated above.)  Using an edge of length 1,
connect each literal vertex to the vertices
representing the clauses in which the literal occurs.

Now a truth assignment induces a wake-up tree of makespan
$3/2+O(\varepsilon\log n)$.
After going through the initial binary tree,
we have $n$ awake robots at the preliminary leaves. For each variable,
pick the node corresponding to the literal in the given truth assignment.
Use $c(x)$ of
the robots close to this literal to wake up all corresponding
clause nodes, and the remaining robot to wake up the
counterpart $\overline{x}$ of $x$.

Conversely, consider a solution of makespan
$3/2+O(\varepsilon\log n)$, and a wakeup path from the root to a
robot at a clause vertex. Clearly, such a path must
contain at least one long edge of length 1, and an odd number of edges
of intermediate length 1/2. Assume that all such paths contain
precisely one edge of length 1, and one of length 1/2.
Consider all the auxiliary clauses of type $(x_i\vee\overline{x_i})$
and their wakeup paths from the root. There are $n$ of these paths,
and $n$ robots within distance $O(\varepsilon\log n)$ from the root.
By the time $1/2+O(\varepsilon\log n)$, for each auxiliary clause,
one awake robot must be within distance $1+O(\varepsilon\log n)$.
This means that at time $1/2+O(\varepsilon\log n)$,
for each of the $n$ variables precisely one of the $n$
robots close to the root must have moved to either vertex $x_i$
or to vertex $\overline{x_i}$, but not both. This means that
the path of robots induces a truth assignment for the variable.
Furthermore, at time $1/2+O(\varepsilon\log n)$
there must be one awake robot within distance 1
of each clause vertex; therefore, all clauses are satisfied
by the induced truth assignment.

This means that there cannot be a solution of makespan
$3/2+O(\varepsilon\log n)$, if there is no satisfying truth assignment.
Furthermore, if there is no solution with only one short and one intermediate
edge on each wakeup path to a clause vertex, any such path in
an optimal solution must have at least two long edges and one
intermediate edge, or at least one long edge and three intermediate
edges. This means that if there is no satisfying truth assignment,
an optimal solution must have makespan at least
$5/2+\Omega(\varepsilon\log n)$.

This concludes the proof.
\end{proof}

\section{Freeze-Tag in Geometric Spaces}
\label{sec:geo}

We now assume that the domain ${\cal D}=\Re^d$ and that the distance,
$d(p_i,p_j)$, between the points $p_i,p_j\in P$ is the Euclidean
distance (or any $L_p$ metric).  In this section, we begin by showing
a constant-factor approximation. We then introduce the notion of
``pseudo-balanced'' awakening trees.  Finally, we show how these two
ideas are combined to yield an efficient PTAS for the geometric
Freeze-Tag Problem.

\subsection{Constant-Factor Approximation Algorithm}

\begin{theorem}
\label{thm:O(1)-approx-plane}
There is an $O(1)$-approximation algorithm, with running time $O(n\log
n)$, for the geometric FTP in any fixed dimension $d$.  The
algorithm yields a wake-up schedule with
makespan $O(diam(R))$, where $diam(R)$ denotes the diameter of
the point set $R$.
\end{theorem}

\begin{proof}
  For each of the points $v\in R$, we consider $K$ sectors defined by
  rays emanating from the points at angles $0,2\pi/K, 2(2\pi/K),
  3(2\pi/K),\ldots$.  Let $u_j(v)$ denote the point (if any) of $R$ in
  the $j$th sector that is closest to $v$; if there are no points of
  $R$ in the $j$th sector of $v$, then $u_j(v)$ is undefined.  We can
  compute the $u_j(v)$ points, for all $j$ and all $v\in R$, in total
  time $O(Kn\log n)$, using standard Voronoi diagram-based
  methods~\cite{c-aaspm-87}.

  We sort the points $u_j(v)$ by distance from $v$; let these points
  be $u_1, u_2,\ldots, u_{K'}$, in sorted order by distance from $v$.
  The wake-up strategy we employ is as follows: Once the robot at $v$
  is unfrozen, it follows the path $v, u_1, u_2,\ldots, u_{K'}$,
  awakening the nearest robot in each of the $K'\leq K$ nonempty
  sectors about it.  (Of course, some of these robots may have
  already been awakened before it gets to their (initial) positions.
  This potentially saves it the effort of going to all of these
  nearby neighbors, allowing for some possible further improvement in
  our constant factor.)

  We now analyze the performance of this algorithm.  Let
  $G_K=(R,E_K)$ be the graph that links each point $v$ to the
  points $u_j$ that are its nearest neighbors in the $K$ sectors about
  $v$.  Such a graph $G_K$ is known as a {\em $\Theta$-graph\/}, for
  $\Theta=2\pi/K$, and is known to be a $t$-spanner for values of
  $K\geq 9$ (e.g., see \cite{e-sts-00}).
  This means that distances in the graph $G_K$ approximate to within a
  constant factor the Euclidean lengths of the edges in the complete
  graph $G$ on~$R$.

  Assume that the robot at point $v_\ell$ is the last one to be
  unfrozen by our algorithm.  What is the path length of the
  ``signal'' (the unfreezing tag) in getting from $v_0$ to $v_\ell$?
  We know that if some point $v$ is reached by the signal by distance
  $t$, then any neighbor, $u_j(v)$, of $v$ in the graph $G_K$ is
  reached by distance $\leq t+\xi$, where $\xi$ is the length of the
  path $v,u_1,u_2,\ldots,u_j$; thus, $\xi\leq (2j-1)\cdot d(v,u_j)\leq
  (2K-1)\cdot d(v,u_j)$.  Thus, the signal will reach $v_\ell$ within
  a distance of at most $(2K-1)$ times the distance from $v_0$ to
  $v_\ell$ in $G_K$.  For constant $K\geq 9$, distances in
  $G_K$ approximate distances in the Euclidean plane, up to a constant
  depending on $K$. This implies that the signal gets to $v_\ell$
  within distance $O(d(v_0,v_\ell))$; 
because $d(s,p_\ell)$ is a lower bound on the optimal makespan, $t^*$,
we have shown that we have an $O(1)$-approximation. 
\end{proof}

\begin{figure}[htb]
\centerline{\psfig{figure=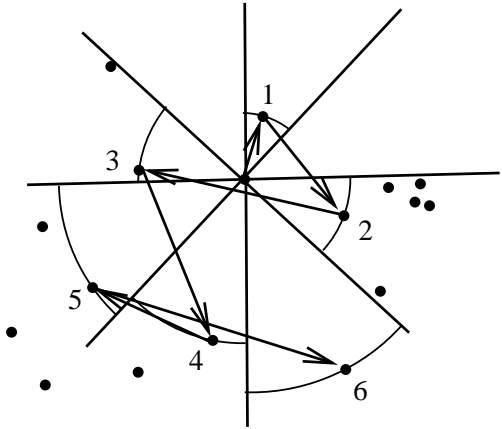,width=0.4\textwidth}}
\caption{An $O(1)$-approximation algorithm for
the geometric FTP in any fixed dimension $d$.  The algorithm generates
a wake-up schedule with makespan $O(diam(R))$.
When a robot at point $p$ first awakens, it awakens the nearest asleep robot
in each of $K$ sectors, in order of increasing distance from
the point $p$.
}
\label{fig:geometric-const-approx}
\end{figure}

\subsection{Pseudo-Balanced Awakening Trees}
\label{sec:pseudo-balanced}

We show how to transform an arbitrary awakening tree
 ${\cal T}$ into an awakening tree ${\cal T}_b $
whose makespan is only marginally longer,
and where no root-to-leaf path travels through too many edges.
This transformation
is critical for the PTAS that follows in the next subsection.

We say that a wake-up tree is {\em pseudo-balanced\/} if each
root-to-leaf path in the tree has $O(\log^2 n)$ nodes.
We have the following theorem:

\begin{theorem}
\label{thm:awakenpolylognodes}
Suppose there exists an awakening tree, ${\cal T}$,
having makespan $t$.  Then, for any $\mu>0$, there exists a
pseudo-balanced awakening tree, ${\cal T}_b$, of makespan $t_b\leq
(1+\mu)t$.
\end{theorem}

\begin{proof}
  First we perform a heavy-path decomposition (e.g., see~\cite{t-dsna-83})
of the tree ${\cal T}$.  For each node $v$, let
  $d(v)$ be the number of descendents.  Consider a node $u$ with
  children $v_1,\ldots,v_k$.  The edge $(u,v_i)$ is {\em heavy\/} if
  $v_i$ has more descendents than any other child of $u$; i.e., 
  $i=\arg\max_j d(v_j)$.
  If $(u,v_j)$ is a {\em light\/} (non-heavy) edge, then at most half of
  $u$'s descendents are $v_j$'s descendents; that is, $d(v_j)\leq
  d(u)/2$.  Thus, in any root-to-leaf path in ${\cal T}$ there are at
  most $\log n$ light edges.  Also, heavy edges form a collection of
  disjoint paths
(because there is one heavy edge from a node to
    one of its children).  We say that a heavy path $\pi'$ is a {\em
    child\/} of heavy path $\pi$ if one end node of $\pi'$ is the child
  of a node in $\pi$.  The heavy-path decomposition forms a {\em balanced
  tree\/} of heavy paths, because any root-to-leaf walk in ${\cal T}$
  visits at most $\log n$ light edges, and therefore at most $\log n$
  heavy paths.

\begin{figure}[htbp]
\begin{center}
\leavevmode
\epsfig{file=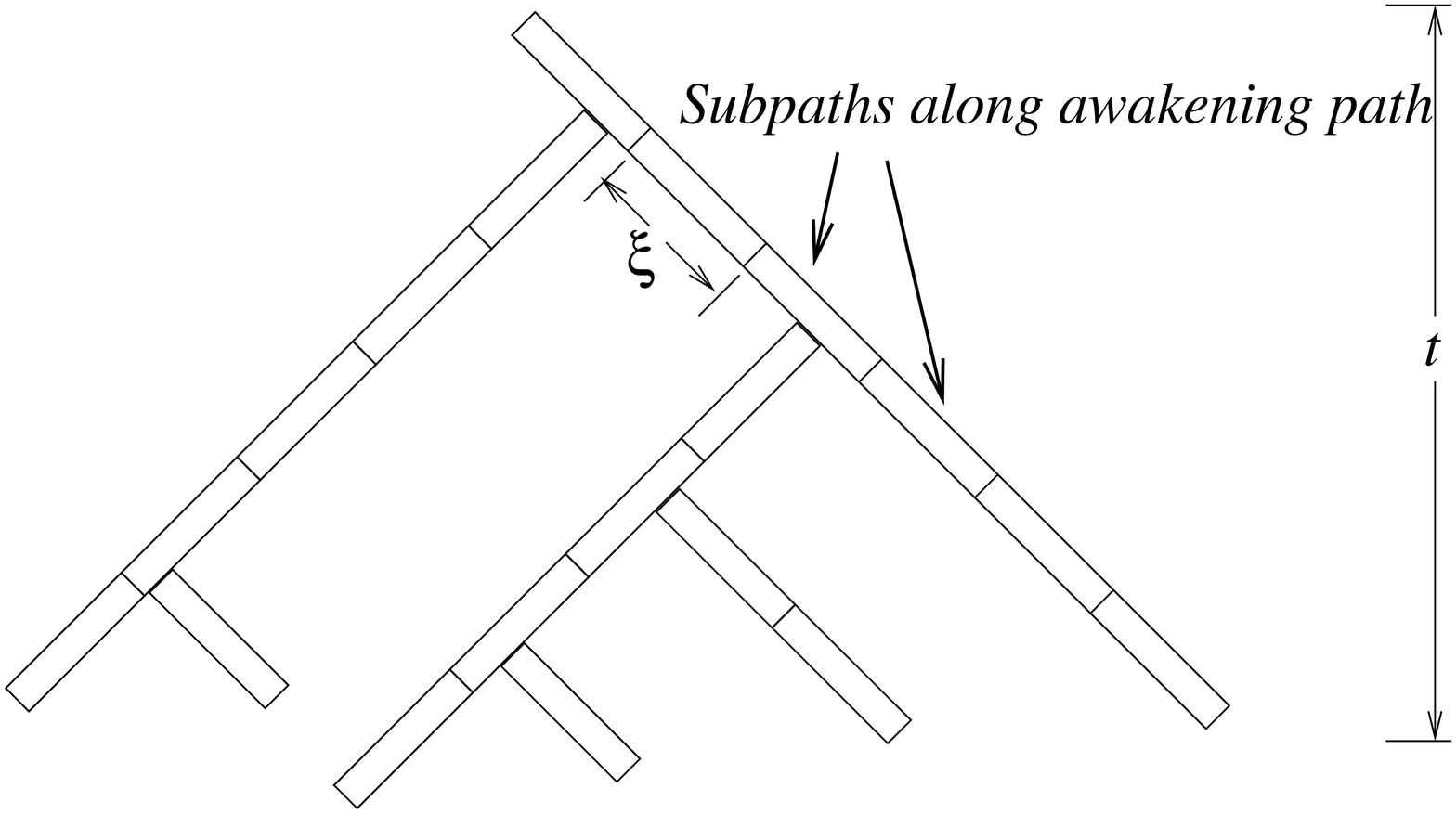,width=0.4\columnwidth}   
\caption{The awakening tree is partitioned into heavy
paths, each of which is partitioned into subpaths
of length $\xi$.}
\label{fig:michael1}
\end{center}
\end{figure}

  We use these heavy paths to refine the description of the
  wake-up tree.  See Figure~\ref{fig:michael1}.
We can assume that in $T$ each heavy path is awakened by one
  robot, the robot that awakens the {\em head\/} of the heavy path
  (node closest to $v_0$)
  and that no robot awakens more than one heavy path.
In this way, a heavy-path decomposition of ${\cal T}$ corresponds to
  an awakening schedule with one robot per path.

Because ${\cal T}$ has makespan $t$, each heavy path has length at most
$t$.  We divide the heavy path into
{\em subpaths\/} of length $\xi=\mu t/(2\log n)$.
Note that on any root-to-leaf path in ${\cal T}$, we visit at
most $O((1+1/\mu)\log n)$ different subpaths.  In the original wake-up
tree, all nodes in one length $\xi$ subpath are awakened by a single
robot.  Thus, by construction, a robot $\delta$ units from the beginning
of the subpath is awakened $\delta$ units after the beginning (head) of the
subpath.  In our modified solution, the robots in a length $\xi$
subpath share in the collective awakening of all the robots in the
subpath.

We guarantee that we can begin awakening one subpath $\xi$ time
units after we began awakening the previous subpath. We further
guarantee that all of the robots are awake and back in their
original asleep positions by
$2\xi$ time units after the first robot in the subpath is
originally awakened.  Thus, a robot $\delta$ units from the
beginning of the subpath is only guaranteed to be awake $2\xi$
units after the robot at the beginning of the subpath is awakened,
which could entail a total delay of $2\xi$ over the original
awakening.

\begin{figure}[htbp]
\begin{center}
\leavevmode
\epsfig{file=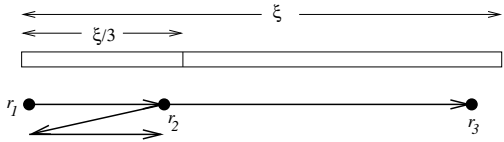,width=0.4\columnwidth}   
\caption{Robot $r_1$ awakens the subproblem $(r_1,r_3)$ by
first awakening $r_2$, the last robot (if any) before
distance $\xi/3$.  Robot $r_2$ is then in charge of awakening $(r_1,r_2)$
before returning to its original position.  Robot $r_1$ then awakens
$(r_2,r_3)$.}
\label{fig:michael2}
\end{center}
\end{figure}

We awaken a subpath as follows; see Figure~\ref{fig:michael2}. 
We consider the subpath to be
oriented from ``left'' (the head, closest to source $v_0$) to
``right''. The first robot $r_1$, at the left end of the subpath,
travels along the subpath until the last (asleep) robot, $r_2$,
before position $\xi/3$, if such a robot $r_2$ exists.  If robot
$r_2$ exists, then $r_2$ is sent leftwards with the responsibility
to awaken all asleep robots in the interval $(r_1,r_2)$, and this
subproblem is solved recursively; thus, $r_2$ is responsible for
initiating the awakening of all robots in the interval
$(r_1,r_2)$, and all robots must return to their initial
positions. If no robot $r_2$ is encountered by $r_1$ before
position $\xi/3$, then we use $r_1$ to solve recursively the
subproblem $(\xi/3,\xi)$.

We continue the strategy until a subproblem's length drops below
$\xi/\log n$ and then resort to a different wake-up strategy. The
responsible robot, $r$, goes to the median robot of the subproblem
and awakens it, and continues in its same direction. The robot it
just awakened goes in the opposite direction and recursively does
the same thing, heading for the median in its subproblem, etc.
Because a segment has at most $n$ robots in it, this strategy takes
time at most $\log n\cdot\xi/\log n$.

Consider a heavy path composed of subpaths of length $\xi$.
Consider any robot at position $\delta$ along the heavy path. The
original wake-up tree will awaken this robot $\delta$ units after
the first robot of the heavy path.  The new solution may awaken
this robot as much as $\delta+2\xi$ time units after the first
robot of the heavy path; one additive delay of $\xi$ is from the
first phase of the awakening and the second additive delay of $\xi$
is from the second phase of the awakening.

Because there are at most $\log n$ heavy paths on any root-to-leaf
walk and there is an accumulated delay of at most $2\xi$ per heavy
path, the total delay on any root-to-leaf path is at most
$2\xi\log n$. Because $\xi=\mu t/\,2\log n$, the accumulated delay in
the makespan is at most $\mu t$.

On any root-to-leaf path in ${\cal T}$ there are at most $O(\log
n)$ subpaths.  Each of these subpaths in our new wake-up tree is
transformed into a wake-up subtree of height $O(\log n)$. Thus, on
any root-to-leaf path in the new wake-up tree there are at most
$O(\log^2 n)$ nodes, and therefore our wake-up tree is
pseudo-balanced.
\end{proof}

\subsection{PTAS}

We give a $(1+\varepsilon)$-approximation algorithm (PTAS) for the
Euclidean (or $L_p$) Freeze-Tag Problem in any fixed dimension.  Our
algorithm runs in nearly linear time, $O(2^{poly(1/\varepsilon)}+ n\log
n)$.  It is important to note that the exponential dependence on
$1/\varepsilon$ appears additively in our time bound, not multiplying
$n\log n$.

\begin{figure}[htb]
\centerline{\psfig{figure=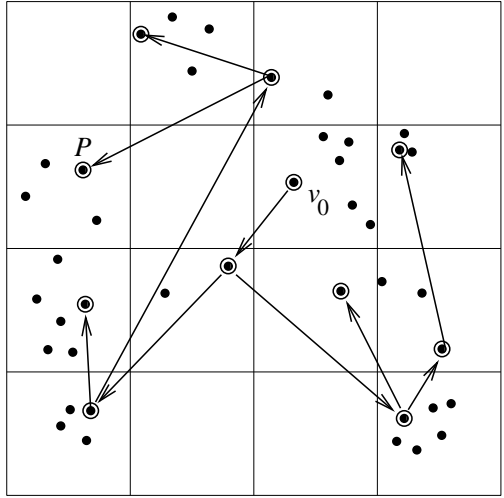,width=0.4\textwidth}}
\caption{PTAS for geometric instances.
Rescale so that all robots lie in a unit square.
Look at $m$-by-$m$ grid of pixels, where $m=O(1/\varepsilon)$.
Consider an enumeration over a special class of wake-up trees on
a set $P$ of representative points, one per occupied pixel.}
\label{fig:geometric-ptas}
\end{figure}

We divide the plane into a constant number of square tiles
or {\em pixels\/}.  Specifically, we rescale the coordinates of the $n$
input points (robots), $R$, so that they lie in the unit square, and we
subdivide the square into an $m$-by-$m$ grid of pixels, each of side length
$1/m$.  (We will select $m$ to be $O(1/\varepsilon)$.)  We say that a
pixel is {\em empty\/} if it contains no robots.

Our algorithm is based on approximately optimizing over a restricted
set of solutions, namely those for which all of the robots within a
pixel are awakened before any robot leaves that pixel.  Note that by
Theorem~\ref{thm:O(1)-approx-plane}, once one robot in a pixel
has been awakened, all of the robots in the pixel can
be awakened within additional time $O(1/m)$,
because this is the diameter of the pixel.

We now describe the algorithm.  We select an arbitrary {\em
  representative\/} point in each nonempty pixel. We pretend that all
robots in the pixel are at this point, and we enumerate over all
possible wake-up trees on the set, $P$, of representative points.  (If
there are $r$ robots in a given pixel, then we only enumerate
wake-up trees whose corresponding out-degree at that pixel is at most
$\min\{m^2-1,r+1\}$.) Because there are only a constant number of
such trees (at most $2^{O(m^2\log m)}$, because $|P|\leq m^2$), this
operation takes time $2^{O(m^2\log m)}$, which is a
  constant independent of $n$.
Recall that a wake-up tree is {\em pseudo-balanced\/} if each
root-to-leaf path in the tree has $O(\log^2 m)$ nodes.
Among those wake-up trees for $P$ that are pseudo-balanced,
we select one, ${\cal T}_b^*(P)$, of minimum makespan, $t_b^*(P)$.
We convert ${\cal T}_b^*(P)$ into a wake-up tree for {\em all\/} of the
input points $R$ by replacing each $p\in P$ with an $O(1)$-approximate
wake-up tree for points of $R$ within $p$'s pixel, according to
Theorem~\ref{thm:O(1)-approx-plane}.  This step takes total time
$O(n\log n)$.  The total running time of the algorithm is therefore
$O(2^{O(m^2\log m)}+n\log n)$.
Correctness is established in the following lemmas.

\begin{lemma}
\label{lem:1}
There is a choice of representative points $P$ such that the makespan of
an optimal wake-up tree of $P$ is at most $t^*(R)$.
\end{lemma}

\begin{proof}
  For each pixel, we select the representative point to be the
  location of the first robot that is awakened in an optimal solution,
  ${\cal T}^*(R)$, for the set of all robots.  Then, for this choice
  of $P$, the spanning subtree of $P$ within ${\cal T}^*(R)$ defines a
  feasible wake-up tree for $P$ of makespan no greater than that of
  ${\cal T}^*(R)$ (namely, $t^*=t^*(R)$).
\end{proof}

\begin{lemma}
\label{lem:3}
For any two choices, $P$ and $P'$, of the set of representative points,
we have $t_b^*(P)\leq t_b^*(P')+O((\log^2 m)/m)$.
\end{lemma}

\begin{proof}
  Pixels have size $O(1/m)$ and there are at most $O(\log^2 m)$
  awakenings in each root-to-leaf path of a pseudo-balanced tree;
  thus, any additional wake-up cost is bounded by $O((\log^2 m) /m)$.
\end{proof}

A similar proof yields:

\begin{lemma}
\label{lem:4}
For any pseudo-balanced wake-up tree of $P$, there exists
a wake-up tree, ${\cal T}(R)$, with makespan $t(R)\leq t_b(P)+O((\log^2 m)/m)$.
\end{lemma}

In summary we have the following result:

\begin{theorem}
\label{thm:ptas-geometry}
There is a PTAS, with running time $O(2^{O(m^2\log m)}+n\log n)$, for
the geometric FTP in any fixed dimension~$d$.
\end{theorem}

\begin{proof}
The time bound was already discussed.
The approximation factor is computed as follows.
By the lemmas above, the makespan, $t$, of the
wake-up tree we compute obeys:
\begin{eqnarray*}
t &\leq& t_b^*(P)+O((\log^2 m)/m)\\
  &\leq& t_b^*(P')+2\cdot O((\log^2 m)/m)\\
  &\leq& t_b(P') + O((\log^2 m)/m)\\
  &\leq& (1+\mu)t^* + O((\log^2 m)/m)\\
  &\leq& t^*\left(1 + \mu + \frac{C\log^2 m}{m}\right)\\
  &\leq& t^*(1+\varepsilon),
\end{eqnarray*}
for appropriate choices of $\mu$ and $m$, depending on $\varepsilon$.
(We also used the fact that $t^*\geq diam(R)\geq 1$.)
\end{proof}

At this point the complexity of the FTP for geometric distances in $\Re^d$
has been unsettled for several years. In fact, this issue is the 
topic of Problem \#35 on the well-known list \cite{TOPP} 
known as ``The Open Problems Project''. We have the following conjecture.

\begin{conj}
\label{con:geo}
The freeze-tag problem is NP-hard for Euclidean or Manhattan distances
in the plane.
\end{conj}

\section{Conclusion}

We have introduced the Freeze-Tag Problem. We have given a number
of algorithmic results for various scenarios. 
For the case of star graphs, we have shown NP-hardness, and analyzed
approximation algorithms, in particular for the case of an identical 
number of robots at each leaf, for which we have given
a simple 7/3 greedy algorithms, and a more complicated PTAS.
We have also shown the existence of constant-factor approximation
methods for general star scenarios, and a 5/3 bound on the 
approximation ratio in general weighted graphs, even for bounded
degree and one robot at each node. Furthermore, we have studied
the Freeze-Tag Problem in geometric spaces, where we showed
the existence of constant-factor approximation algorithms,
including a PTAS.

Obviously, there
is a considerable number of open problems that deserve further
study:
\begin{enumerate}
\item Is there a lower bound on the approximability of the
  Freeze-Tag Problem on tree metrics?
\item Is there an $o(\log n)$-approximation algorithm
for the Freeze-Tag Problem in general weighted graphs?
\item Is the FTP in low-dimensional geometric spaces
NP-hard?
\item Is there a PTAS for the FTP in trees with different numbers
of robots at each leaf?
\item Is there an $o(\log n)$- (or, ideally, an $O(1)$-)
approximation algorithm for the FTP for points in a
polygon, where distances are measured according
to the length of a shortest path in the polygon?
Such an algorithm would apply also to the FTP in general trees.
\item Can our results be extended to the case of several sources?
\item In a geometric scenario, how does the problem change if a robot
only has to get ``close''
  to another robot (say, within distance 1) in order to unfreeze it?
\end{enumerate}

It is also of interest to consider more game-theoretic aspects related
to freeze tag, like considering algorithmic issues arising
from the full game
of freeze-tag in the presence of an adversary. This is somewhat related to
the {\em Competing Salesman Problem} \cite{fffs-tspc-04}, where two salesmen
travel in a graph and try to visit vertices before the opponent
does.

As described in the
introduction, there are many other related questions, and we do expect
many more interesting results arising from this research.

\subsection*{Acknowledgments}

We thank Doug Gage for discussions motivating this research.
We are also grateful to Regina Estkowski, Tien-Ruey Hsiang, Nenad Jovanovic,
David Payton, and Marcelo Sztainberg for many helpful discussions.
We thank Asaf Levin for pointing out the lower bound in Section~4.1 and for
alerting us to the omission of the assumption of locally bounded edge
weights in Theorem~\ref{th:approx.deg} in the conference draft of this
paper. We thank Gerhard Trippen and other anonymous referees for several
useful comments that helped to improve the overall presentation of this
paper.

Esther Arkin acknowledges support from HRL Laboratories, the National
Science Foundation (CCR-0098172, CCF-0431030), and Sandia National
Laboratories.  Michael Bender acknowledges support from HRL
Laboratories, the National Science Foundation (EIA-0112849), and
Sandia National Laboratories.  Work by S\'andor Fekete was performed
in part while visiting Stony Brook, with partial support by the DFG
Focus Program 1126, ``Algorithmic Aspects of Large and Complex
Networks,'' grant no.~Fe~407/8-1.  Joseph Mitchell acknowledges
support from HRL Laboratories, Honda Fundamental Research Labs, Metron
Aviation, NASA Ames Research Center (NAG2-1325, NAG2-1620), the
National Science Foundation (CCR-0098172, ACI-0328930, CCF-0431030),
and the US-Israel Binational Science Foundation (Grant 2000160).
Martin Skutella is supported in part by the EU Thematic Network
APPOL~I+II, ``Approximation and Online Algorithms,'' IST-1999-14084
and IST-2001-30012, and by the DFG Focus Program 1126, ``Algorithmic
Aspects of Large and Complex Networks,'' grant no.~SK~58/4-1.

\bibliographystyle{abbrv}
\bibliography{refs}

\end{document}